\documentclass[aps,physrev,preprint,superscriptaddress,amsmath,amssymb,]{revtex4-2}
\usepackage[T1]{fontenc}
\usepackage{graphicx}

\renewcommand{\phi}{\varphi}
\usepackage{upgreek} 

\begin{document}

\title{Reservoir neuromorphic computing based on spin-orbit coupling in an organic crystal resonator}

\author{Teng Long}
\altaffiliation{These authors contributed equally to this work.}
\affiliation{Beijing Key Laboratory for Optical Materials and
Photonic Devices, Department of Chemistry, Capital Normal
University, Beijing 100048, People’s Republic of China}

\author{Yibo Deng}
\altaffiliation{These authors contributed equally to this work.}
\affiliation{Beijing Key Laboratory for Optical Materials and
Photonic Devices, Department of Chemistry, Capital Normal
University, Beijing 100048, People’s Republic of China}

\author{Xuekai Ma}
\affiliation{Department of Physics and Center for
Optoelectronics and Photonics Paderborn (CeOPP),
Paderborn University, 33098 Paderborn, Germany}

\author{Chunling Gu}
\affiliation{College of New Materials and Chemical Engineering, Beijing institute of petrochemical technology, Beijing, 102617, People’s Republic of China}

\author{Guillaume Malpuech}
\affiliation{Universit\'e Clermont Auvergne, Clermont Auvergne INP, CNRS, Institut Pascal, F-63000 Clermont-Ferrand, France}

\author{Qing Liao}
\email{liaoqing@cnu.edu.cn}
\affiliation{Beijing Key Laboratory for Optical Materials and
Photonic Devices, Department of Chemistry, Capital Normal
University, Beijing 100048, People’s Republic of China}
\affiliation{School of Physics and Materials Science, Guangzhou University,
Guang-zhou 510006, People’s Republic of China}

\author{Hongbing Fu}
\email{hbfu@cnu.edu.cn}
\affiliation{Beijing Key Laboratory for Optical Materials and
Photonic Devices, Department of Chemistry, Capital Normal
University, Beijing 100048, People’s Republic of China}
\affiliation{College of New Materials and Chemical Engineering, Beijing institute of petrochemical technology, Beijing, 102617, People’s Republic of China}

\author{Dmitry Solnyshkov}
\affiliation{Universit\'e Clermont Auvergne, Clermont Auvergne INP, CNRS, Institut Pascal, F-63000 Clermont-Ferrand, France}
\affiliation{Institut Universitaire de France (IUF), 75231 Paris, France}

\begin{abstract}
Neuromorphic computing is at the basis of the recent progress in artificial intelligence. But the progress is accompanied with increasing demands in computational resources and power supply.
Reservoir neuromorphic computing uses a non-linear physical system to replace a part of a large neural network. The advantages can include reduced power consumption and faster learning.
We show that the interference in an organic crystal waveguide resonator leads to efficient separation of optical patterns, allowing a significant reduction of the size of the neural network  and an  acceleration of the learning process. For more complex symbols, extending the reservoir output dimension thanks to spin-orbit coupling, we achieve a 10-times reduction of the network size and a 3-fold speedup. Our work suggests a general path for the performance improvement of photonic reservoir computing systems.
\end{abstract}

\maketitle

\vspace{3 cm}

Artificial intelligence is rapidly penetrating into all aspects of our life. As the amount of neural networks is increasing, it becomes more and more important to improve their key characteristics~\cite{markovic2020physics}: to shorten their training time, reduce the power consumption during the training and their use, and to increase their adaptivity. Neuromorphic computing is therefore a subject of intense fundamental and applied research, that constantly explores new physical routes in photonics~\cite{lin2018all,wetzstein2020inference,shastri2021photonics,matuszewski2024role}, including strongly-coupled light-matter systems~\cite{kavokin2022polariton,opala2022training,tyszka2023leaky,dini2024nonlinear, sedov2025polariton}.

One of the approaches which has drawn a strong attention of physicists is the so-called reservoir neuromorphic computing~\cite{jaeger2004harnessing} with analog implementation, where a part of a big neural network is replaced by a non-linear physical system~\cite{Lukos2009} (Fig.~\ref{fig1}a,b). The choice of the physical system is crucial, as it must be able to separate different input signals while keeping similar signals clustered in the multidimensional output space~\cite{tanaka2019recent}. A properly chosen physical system allows improving key characteristics of the whole neuromorphic system: as the depth of the trained network is reduced, the training is much faster and less power-consuming.

Physical implementations of reservoirs are very diverse, including fluids~\cite{gao2023optofluidic,marcucci2023new}, solid-state~\cite{farronato2023reservoir,yaremkevich2023chip}, and optical systems.
Photonic implementations of reservoir neuromorphic computing have blossomed since 2008, with a lot of proposals~\cite{Vandoorne08,tanaka2019recent} and experimental  implementations~\cite{vandoorne2014experimental,brunner2015reconfigurable}.
Some implementations were based on the strong interactions of exciton-polaritons~\cite{ballarini2020polaritonic} and the corresponding non-linearities of the refractive index~\cite{gan2025ultrafast}, with potential quantum computing outlooks~\cite{ghosh2021quantum,krisnanda2025experimental}. However, the simplest realization of non-linearity, used since the first works on reservoir computing, such as Ref.~\cite{fernando2003pattern}, studying water waves,
is the interference, where the sum of $N$ equal amplitudes $A$ (for example) can give an intensity $I$ between $0$ and $N^2 |A|^2$, depending on the relative phase (in general, $I=|\sum_{n=1}^N A_n|^2$). Another important physical effect allowing to promote the dimensionality of the output space of the reservoir (one of its most important parameters~\cite{stenning2024neuromorphic}) is the spin-orbit coupling, already used very successfully for reservoir computing with electrons~\cite{torrejon2017neuromorphic}.  In photonic systems, the most widespread type of spin-orbit coupling~\cite{bliokh2008geometrodynamics} is the splitting between transverse-electric and transverse-magnetic modes (the TE-TM splitting, shown in Fig.~\ref{fig1}(c) combined with birefringence). It is widely used in topological photonics~\cite{Haldane2008}. In planar structures it leads to the optical spin Hall effect~\cite{Kavokin2005} (OSHE), observed in cavities~\cite{leyder2007observation} and even in simplest organic resonators~\cite{ren2025optical}. However, the photonic spin-orbit coupling has not been used for photonic reservoir computing so far.

In this work, we present a simple, cheap and easy to fabricate implementation of a physical reservoir: a 2-dimensional (2D) hexagonal photonic resonator based on an organic crystal waveguide possessing interference-induced non-linearities and spin-orbit coupling~\cite{deng2024spin}, shown schematically in Fig.~\ref{fig1}(e). We demonstrate experimentally that this reservoir provides an efficient separation of a set of 10 simple symbols in a 3-dimensional space of output intensities, providing a 30-times increase of the learning speed. For more complex MNIST symbols, the OSHE allows extending the output space by using the polarization degree of freedom, and the final network size is reduced by a factor 10, providing a 3-fold speedup while maintaining prediction accuracy. The waveguide resonator configuration is favorable for integrated photonics.

\section*{The sample and the setup}

\begin{figure}
    \centering
    \includegraphics[width=0.95\linewidth]{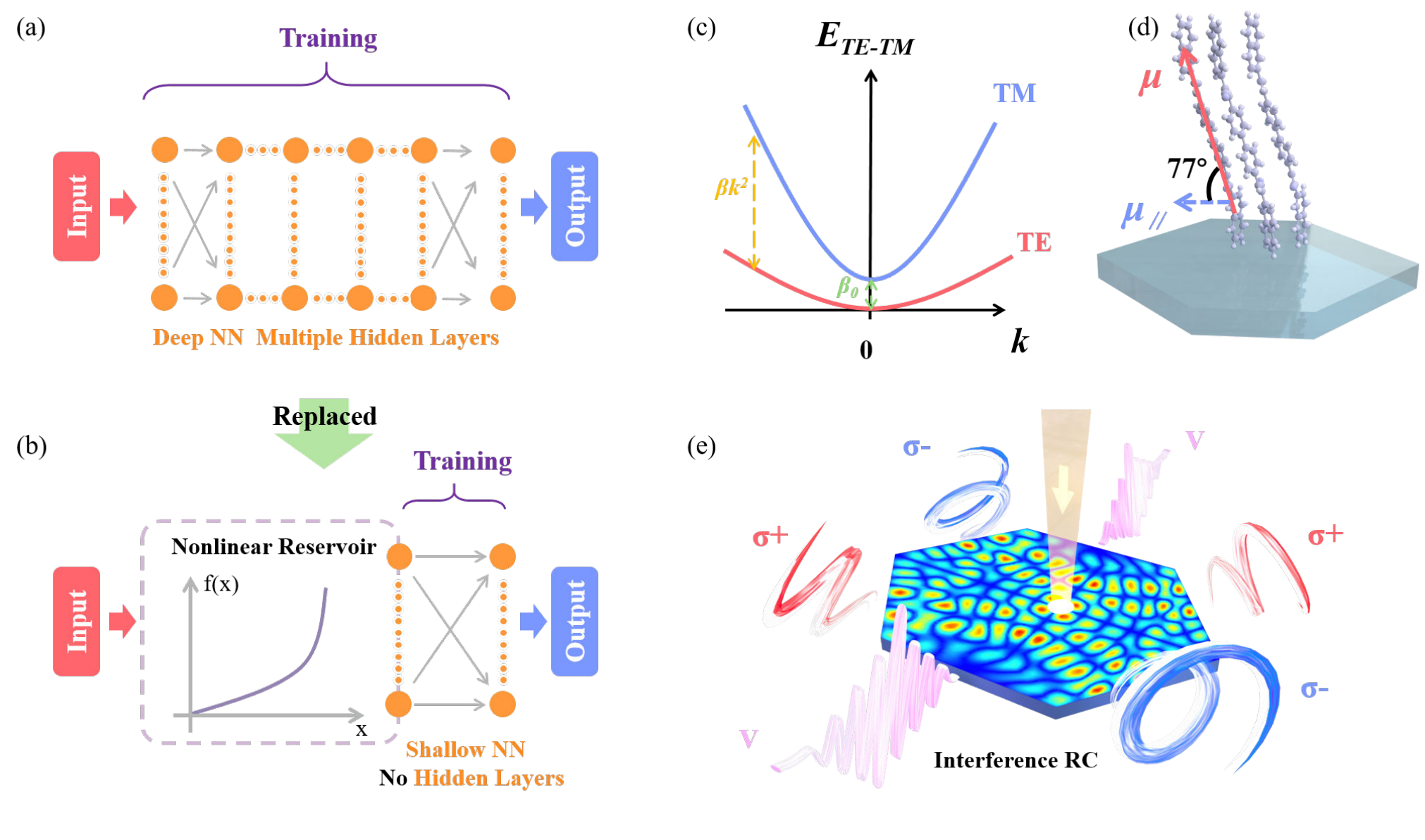}
    \caption{\textbf{Reservoir neuromorphic computing with a spin-orbit-coupled optical resonator.} a) A multi-layer neural network. The training involves adjusting weights in all layers. b) A shallow neural network using a nonlinear physical system as a reservoir. The training only involves the output layer. c) Photonic spin-orbit coupling: the dispersions of the TE and TM modes in a planar resonator.(Experimental measurements in Figure S5) d) The direction of the dipole moment in BPDBNA crystals e) A scheme of a hexagonal optical resonator with the output signal collected from the sides. }
    \label{fig1}
\end{figure}

We utilized (2Z,2’Z)-3,3-([1,1’-biphenyl]-4,4’diyl)bis (2-(naphthalen-2-yl)acrylonitrile) (BPDBNA) 2D hexagonal microcrystals (Fig.~\ref{fig1}d) prepared via the physical vapor deposition method (more details see Method). The BPDBNA molecule exhibits a rod-like structure with a distyrylbenzene core bridged by two naphthalene rings at its termini, ensuring excellent~\cite{deng2024spin} optoelectronic properties (see Figure S1). For such rod-like molecules, the transition dipole moment ($\mu$) is aligned along the molecular long axis, as illustrated in Fig.~\ref{fig1}d. Through precise molecular alignment engineering, the $\mu$ forms a 77° angle with the crystal surface, which not only provides a strong birefringence but also facilitates the coupling of linearly polarized emission perpendicular to $\mu$ into the organic crystal waveguide.

We performed the optical measurements on the resulting BPDBNA hexagonal crystals using photoluminescence (PL) imaging. We note that the hexagon is not regular (see Figure S3), because the BPDBNA molecules crystallizes with a space group symmetry $P2_1/n$F (monoclinic, Table S1).
As shown schematically in Figure~\ref{fig1}(e), a 405-nm continuous-wave (CW) laser is focused into a $\approx 2$ $\mu$m spot at the center of an isolated hexagonal crystal. This excitation generated a strong optical field (photoluminescence - PL) propagating via the optical modes of the resonator. The modes are radiative, but their lifetime is relatively long, which is why the externally observed emission is localized at the crystal edges, which scatter the light from the modes towards the observer. 

In order to demonstrate the strong spin-orbit coupling present in our resonator (Figure S6), the polarization distribution was characterized by measuring Stokes parameters using a quarter-wave plate and linear polarizer in the optical path (Figure S2). We note that we have already used this spin-orbit coupling in BPDBNA resonators to generate circular-polarized electroluminescence~\cite{deng2024spin}.

\section*{Separation, clustering and accelerated learning}

\begin{figure}
    \centering
    \includegraphics[width=0.95\linewidth]{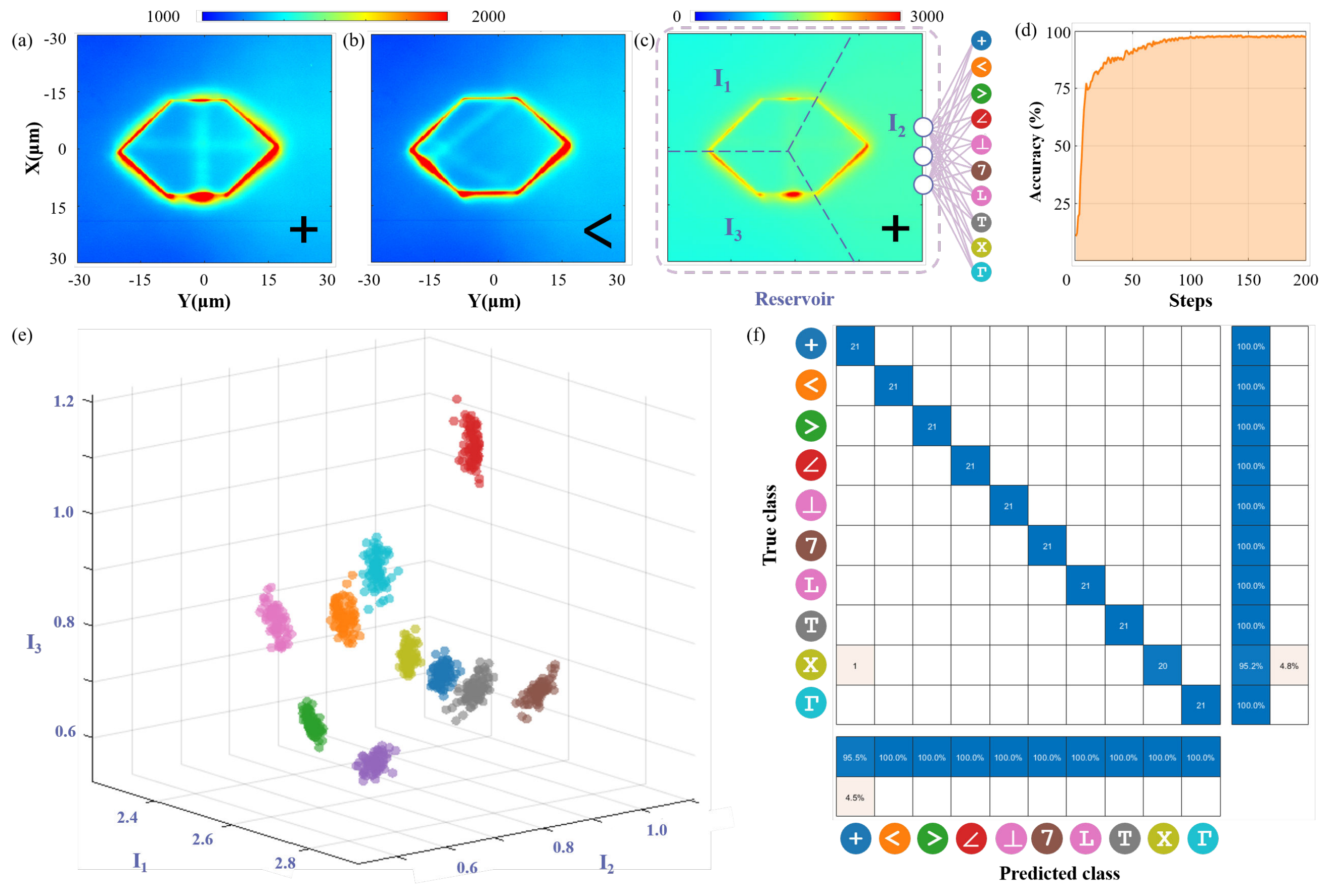}
    \caption{\textbf{Efficiency of experimentally implemented reservoir neuromorphic computing.} a,b) The input symbols on the surface of the resonator; c) The scheme of the system: the reservoir + a $3\times 10$ shallow network. One of the input symbols with the definition of the intensity integration sectors  ($I_1$, $I_2$, $I_3$); d) The rapid increase of the prediction accuracy during training with reservoir. e) The separation and clustering properties: the output signal in the 3D space spanned by $I_1$, $I_2$, $I_3$ for each of the 10 symbols indicated by the colorbar.  f) The confusion matrix plot demonstrating a very high final accuracy.}
    \label{fig2}
\end{figure}

Employing a combination of two strip beams for vertical excitation, we create 10 different symbols ($+$, $<$, $>$, $\angle$, $\perp$, $7$, $\mathrm{L}$, $\mathrm{T}$, $\mathrm{X}$, $\Gamma$) on the surface of the hexagonal microcrystal (see Figure S8). Two examples are shown in Fig.~\ref{fig2}(a,b). The cloud of carriers created by the pump becomes a source of waves propagating over the photonic modes. While the PL emission in general exhibits a broad spectrum which usually leads to the smearing out of the interference effects, in our case the emission detected outside the sample from the  modes exhibits quite a narrow spectrum (see Figure S7), with a linewidth of approximately 30~meV. Because of this relatively narrow linewidth involving a limited number of quantized modes of the resonator, the interference is not strongly smeared out, and the resulting intensity distribution remains sufficiently nonlinear to allow using the hexagonal crystal as a reservoir. Moreover, the possibility to excite a superposition of a sufficient number of the eigenmodes is absolutely required for the reservoir operation, otherwise all input patterns would give the same output pattern (that of the single excited eigenmode).


The signal from the physical reservoir can have a very large dimensionality. In order to gain in efficiency of reservoir neuromorphic computing, this dimensionality needs to be reduced, which is usually achieved by integration. The smaller is the output dimensionality, the faster and simpler becomes the training of the reduced network. However, the integrated output signal must still provide the separability property: different inputs should not be overlapping in the output space.

In our case, the output dimension can be reduced down to 3 by integrating the total emission intensity in the 3 sectors of the hexagon, with each sector containing 2 of the 6 edges, as shown in Fig.~\ref{fig2}(c). The separability property is demonstrated by Fig.~\ref{fig2}(e), showing the points corresponding to different input signals by different colors in the 3D space spanned by the three integrated intensities $I_1$, $I_2$, $I_3$. Each point corresponds to a different realization of the experiment, and the random variation of the input signal is clearly visible. Yet, the points of the same color are clustered in 10 non-overlapping "clouds", separated by sufficiently large empty spaces. It means that for any given point, one can immediately tell to which cluster it belongs, and therefore to determine the chosen symbol.

This means that an extremely simple neural network $3\times 10$ connected to the output of the reservoir as shown in Fig.~\ref{fig2}(c) can be trained very rapidly to recognize the 10 input symbols from the output of the reservoir. The evolution of the prediction accuracy is shown in Fig.~\ref{fig2}(d): it increases very rapidly in only 80 steps of the conjugate gradient method. The associated confusion matrix is plotted in Fig.~\ref{fig2}(f). The final accuracy for this number of steps is 99.5\%.

To determine the increase of the performance thanks to the use of reservoir (Fig.~\ref{fig1}(a,b)), we compare the number of steps needed to train a reduced $3\times 10$ network with a physical reservoir, with a full-sized network whose input dimensions are determined by the resolution of the input CCD camera. As a measure of efficiency, we use the training time (in seconds) required to achieve the same accuracy on a high-performance workstation. We observe approximately a 30-times reduction of the training time in the reservoir-based configuration, which demonstrates the remarkable efficiency of this approach.

\section*{Optical spin Hall effect for the MNIST dataset}

The MNIST dataset, containing 10 handwritten digits, is often used for testing the performance of neural networks~\cite{zhou2021large,liao2025hetero}, including photonic reservoir networks~\cite{spagnolo2022experimental}. These symbols are more complex than the 10 symbols used in the previous section, and they require a higher dimensionality of the output of the physical reservoir. Even considering the 6 edges of the hexagon or splitting each of them into 2 (which gives a 12-dimensional output signal) is not enough: our results show that the accuracy drops down to about 75\% in this case. Our key idea is therefore to extend the dimensions of the output signal by accounting for the polarization effects, known to play an important role in photonic systems.

Due to the anisotropic properties of the BPDBNA molecule, the PL is mostly linearly polarized at the source (because one of the transitions is much stronger than the other), and therefore it excites a direction-dependent superposition of TE and TM modes, allowing one to expect the oscillations between these modes during the beam propagation -- the OSHE, already observed in many systems~\cite{leyder2007observation,ren2025optical}. The spin-orbit coupling effects are particularly strong in organic systems, due to their strong anisotropy~\cite{ren2021nontrivial}.

\begin{figure}
    \centering
    \includegraphics[width=1.00\linewidth]{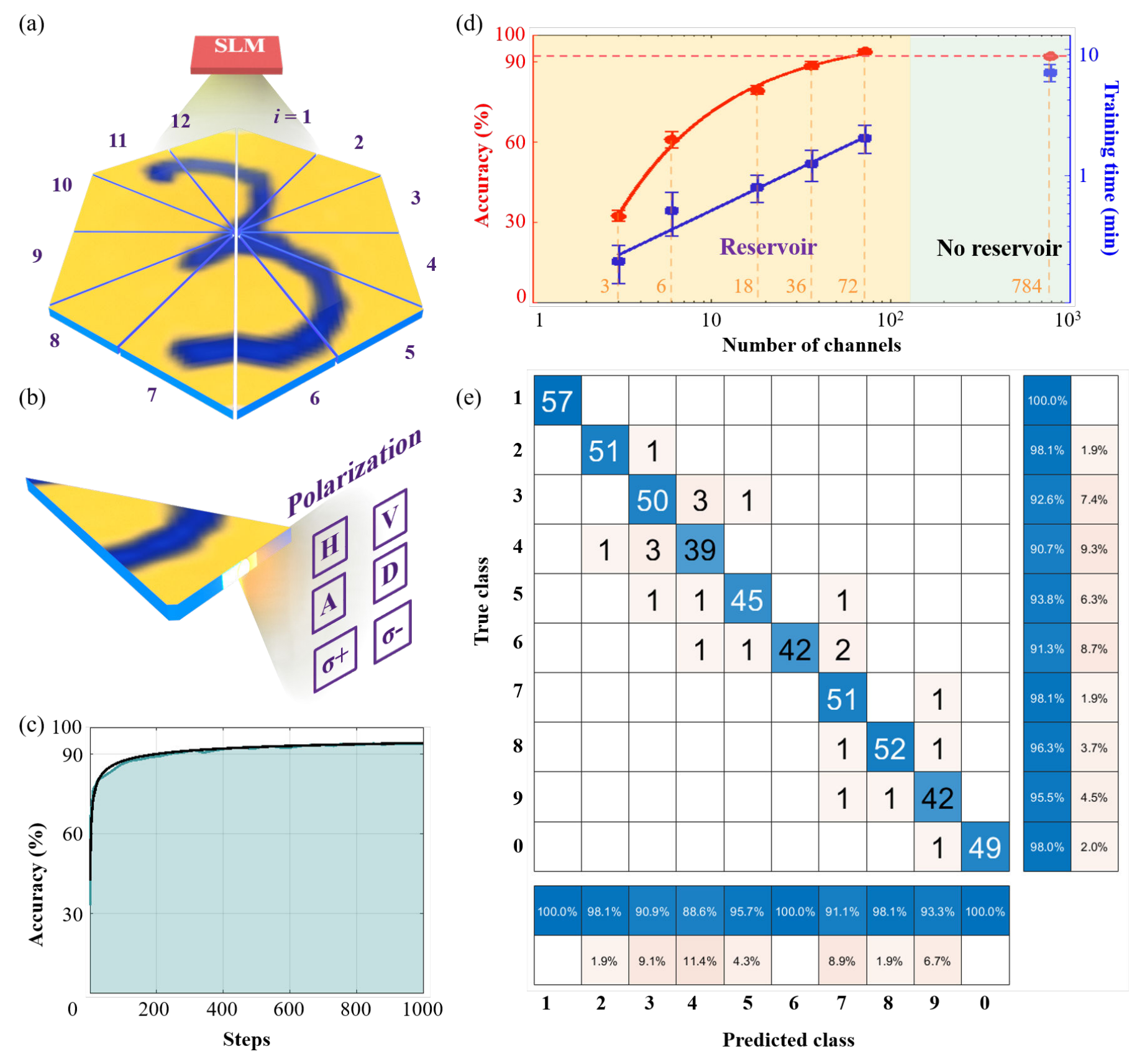}
    \caption{\textbf{Reservoir computing applied to the MNIST dataset.} a) The total intensity distribution created by one of the MNIST symbols in the resonator, with the sectors indicating the intensity integration regions from the edges. b) One of the sectors of the sample, with the emission from the edge decomposed into 6 polarization channels. c) The evolution of accuracy of the trained $72\times 10$ network with the number of steps used in training (solid black line - power law fit). 
    d) The accuracy (left axis, red) and the training time (right axis, blue) as functions of the number of channels in the reservoir. The case without reservoir (784 bit image) is shown on the right as a reference for comparison. Points with error bars (standard deviation) correspond to the experiment, solid lines - power law fits. 
    e) The confusion matrix, corresponding to an overall accuracy of 95.6\% for this particular network. }
    \label{fig3}
\end{figure}

Using an SLM, we have projected 5000 symbols (500 for each digit) from the MNIST database on the sample. Non-resonant excitation light (405~nm CW laser) passes through the SLM and is vertically directed onto the center of the sample (slightly differing in size from the previous sample). By establishing an integrated system between the SLM and CCD using LabVIEW, automatic data acquisition of the dataset was achieved (see Figure S9). We note that the analyzed intensity is not just the MNIST symbol, but an interference pattern created by this symbol as its source. We have measured the intensities of 6 polarization components (right- and left-circular, horizontal, vertical, diagonal, anti-diagonal), which are spatially integrated over 12 sectors along the edges (6 edges split in 2, see Fig.~\ref{fig3}(a,b)), giving the maximal output dimension of 72. In order to analyze the efficiency of reservoir computing as a function of the output dimension, we have also considered smaller dimensions: 3 (total intensity, three 120-degree sectors), 6 (total intensity, six 60-degree sectors), 18 (6 polarizations in 3 sectors), and 36 (6 polarizations, 6 sectors).

Figure~\ref{fig3}(c) shows the increase of the accuracy as a function of the number of training steps for the main case of interest (72 channels). Each point of this figure is obtained by averaging over 20 random initializations of the neural network and random choices of the 90\%/10\% training/test symbols, respectively. The accuracy quickly increases above 80\% and reaches 90\% for 250 steps, after which it slowly grows and saturates at $93.6\pm 1.1$ percent for 1000 steps. This is comparable to the accuracy achieved in a high-dimensional in-sensor reservoir network trained using the whole 60000 set~\cite{jang2024high}.

Figure~\ref{fig3}(d) presents our key results, demonstrating the comparison between  reservoir networks with different reservoir output dimension and a shallow network without reservoir, used as a reference. The red curve demonstrates how the maximal accuracy improves with the number of channels, whereas the blue curve shows the associated training time.

We observe two emerging power laws in the scaling of the experimental results: the error rate decreases with a scaling exponent of $-0.32\pm 0.01$ with the number of steps (for 72 channels) and with a scaling exponent of $-0.71\pm 0.03$ with the number of channels. The training time increases with a scaling exponent of $0.68\pm 0.06$ with the number of channels.
The former of these exponents is in agreement with a recent analytical result of $-1/3$~\cite{ren2025emergence}, while the latter shows the possible efficiency gain associated with reservoir computing, especially for strongly reduced reservoir output dimensions.

Training the resulting $72\times 10$ neural network to treat the output of the physical reservoir (Fig.~\ref{fig3}(e)) allows obtaining a level of accuracy of 93.6\% (for a training subset of 4500 digits), higher than for the shallowest possible network $784\times 10$ (92\%), whose size is determined by the resolution of the MNIST dataset itself (that is, without the physical reservoir). We therefore gain at least a factor 10 on the size of the network by using the physical reservoir, while improving the accuracy. This size reduction provides a 3-fold speedup of the training process, performed on a high-performance workstation GPU using the torch python library.

\section*{Conclusions}

We have demonstrated experimentally that a physical reservoir based on a photonic hexagonal resonator grown from an organic polymer crystal BPDBNA can replace a part of a neural network, allowing to reduce its depth and thus to improve the speed of its training by a factor 30. The nonlinearity is provided by the interference of the photonic modes. We show the separability and clustering of the symbols in the space of reduced dimensions. Applying this approach to the MNIST digit dataset allows one to reduce the network size by a factor 10 and to increase the training speed by a factor 3 without the loss of accuracy thanks to the polarization degree of freedom and the associated OSHE.

The resonator used in our work is particularly simple in fabrication: it does not require any mirrors and consists of a single organic crystal. The reflection provided by the change of the refractive index is sufficient to observe the quantized modes and their interference. While in our experiment we collect the emission from the edges of the sample with an objective located at a large distance (out of plane), in applications the structure can be fully integrated on a chip, by connecting output waveguides to the edges of the resonator. This will also strongly reduce the scattering losses and improve the efficiency of the device. The PL lifetime of our organic crystals is about 5~ns~\cite{deng2024spin}, meaning that the photonic reservoir will not be a bottleneck during the training procedure. The optical power consumption can be estimated as 1.5~pJ per symbol, comparable with that of the most power-efficient reservoir approaches~\cite{yaremkevich2023chip}.
Once the network is trained, symbol recognition can be performed at sub-microsecond timescales.

Our work presents an important example of flexibility offered by a physical reservoir: the number of the output signals can be selected at will as a function of complexity of the input signal. This choice provides an important balance between the speedup and the accuracy of the recognition. The spin-orbit coupling, present in most photonic systems, provides a natural way of controlling the output dimensionality of the reservoir.

\section{Methods}
\textbf{Sample preparation}
BPDBNA single crystals were fabricated using a facile physical vapor deposition (PVD) method. Typically, a quartz boat carrying 10~mg of BPDBNA was placed in the center of a quartz tube which was inserted into a horizontal tube furnace. A continuous flow of cooling water inside the cover caps was used to achieve a temperature gradient over the entire length of the tube. To prevent oxidation of BPDBNA, Ar was used as inert gas during the PVD process (flowrate: 45 sccm min$^{-1}$). The pre-prepared hydrophobic substrates were placed on the downstream side of the argon flow for product collection and the furnace was heated to the sublimation temperature of BPDBNA (at temperature region of $\approx 280^\circ$C), upon which it was physically deposited onto the pre-prepared hydrophobic substrates for 10 hours.
In our case, the chosen hexagonal sheets have smooth surface and uniform morphology with typical width of around 24.7~$\mu$m, length of 36.7~$\mu$m and thickness of 2.5~$\mu$m of hexagonal resonator shown in Figure S3(c).

\textbf{Optical characterization}
The momentum space and two-dimensional real space spectra were acquired using a custom-built angle-resolved micro-angle-resolved spectroscopy system.(see Supplementary Figure S2). The spectrometer equipped with a 300 lines/mm grating and a liquid nitrogen-cooled charge-coupled device (CCD) with a resolution of 400 × 1340 pixels. For non-resonant excitation, a 405 nm CW laser (CNI MDL-H-405) was employed. The GAEA-2.1 Phase Only LCOS-SLM was employed to generate the patterns of the MINST dataset.

\begin{acknowledgments}
We acknowledge useful discussions with A. Scherbakov.
This work was supported by the European Union’s Horizon 2020 program, through a FET Open research and innovation action under the grant agreements No. 964770 (TopoLight) and EU H2020 MSCA-ITN project under grant agreement No. 956071 (AppQInfo). Additional support was provided by the ANR Labex GaNext (ANR-11-LABX-0014), the ANR program "Investissements d'Avenir" through the IDEX-ISITE initiative 16-IDEX-0001 (CAP 20-25), the ANR project MoirePlusPlus (ANR-23-CE09-0033), and the ANR project "NEWAVE" (ANR-21-CE24-0019). We are grateful to the Mésocentre Clermont-Auvergne of the Université Clermont Auvergne for providing help, computing and storage resources.
\end{acknowledgments}

\bibliography{references}

\renewcommand{\thefigure}{S\arabic{figure}}
\setcounter{figure}{0}
\renewcommand{\theequation}{S\arabic{equation}}
\setcounter{equation}{0}

\section{Supplementary Information}

In this supplementary information, we provide additional details of the experiments together with additional results, both theoretical and experimental.

\begin{figure}
    \centering
    \includegraphics[width=0.95\linewidth]{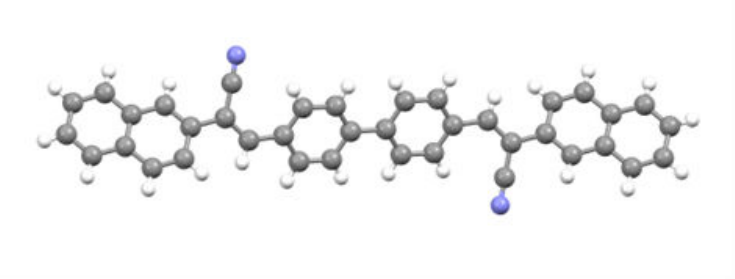}
    \caption{The structure of the (2Z,2’Z)-3,3-([1,1’-biphenyl]-4,4’diyl)bis(2-(naphthalen-2-yl)acrylonitrile) (BPDBNA) molecule. The small balls represent carbon atoms (gray), hydrogen atoms (white), and nitrogen atoms (purple), respectively.}
    \label{figmolscheme}
\end{figure}

\section{The angle-resolved spectroscopy characterization}
The angle-resolved spectroscopy was performed at room temperature by the Fourier imaging using a 100× objective lens of a NA = 0.95, corresponding to a range of collection angle of ±60° (Fig. S2). An incident white light from a Halogen lamp with the wavelength range of 400-700 nm was focused on the area of the microcavity containing a BPDBNA crystal. The angular distribution of the reflected light was located at the back focal plane of the objective lens. Lenses L1-L4 formed a confocal imaging system together with the objective lens, by which the k-space light distribution was first imaged at the right focal plane of L2 through the lens group of L1 and L2, and then further imaged, through the lens group of L3 and L4, at the right focal plane of L4 on the entrance slit of a spectrometer equipped with a liquid-nitrogen-cooled CCD. The use of four lenses here provided flexibility for adjusting the magnification of the final image and efficient light collection. By removing the L3 and opening the slit on the spectrograph, CCD imaging was performed to obtain a two-dimensional (2D) R-space image. 
In order to investigate the polarization properties, we placed a linear polarizer, a half-wave plate and a quarter-wave plate in front of spectrometer to obtain the polarization state of each pixel of the k-space images in the horizontal - vertical (H and V), diagonal - anti-diagonal (D and A) and circular ($\sigma^+$ and $\sigma^-$) basis~\cite{Dufferwiel2015,manni2013hyperbolic}. Thereby, one can calculate the Stokes vector through the equations
$S1=(I_H-I_V)/(I_H+I_V)$, 
$S2=(I_D-I_A)/(I_D+I_A)$,
$S3=(I_{\sigma^+}-I_{\sigma^-})/(I_{\sigma^+}+I_{\sigma^-})$.

\begin{figure}
    \centering
    \includegraphics[width=1\linewidth]{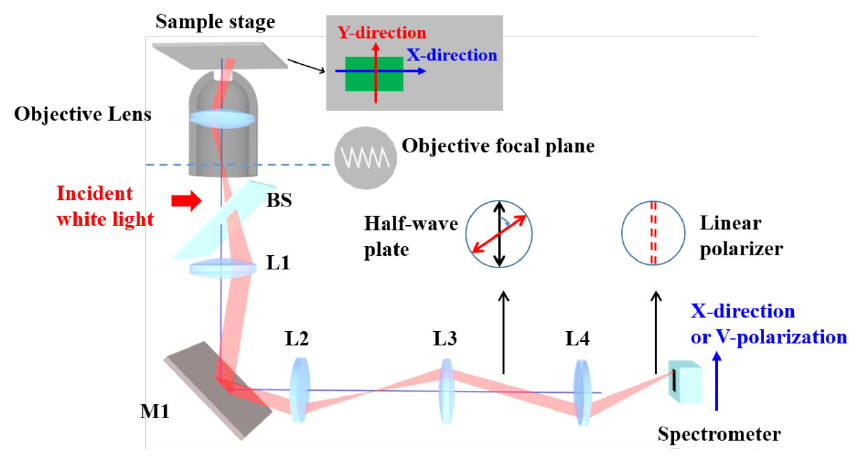}
    \caption{Experimental setup allowing to obtain polarization-resolved complete state tomography. BS: beam splitter; L1-L4: lenses; M1: mirror. The red beam traces the optical path of the reflected light from the sample at a given angle.}
    \label{figsetscheme}
\end{figure}

\begin{figure}
    \centering
    \includegraphics[width=0.8\linewidth]{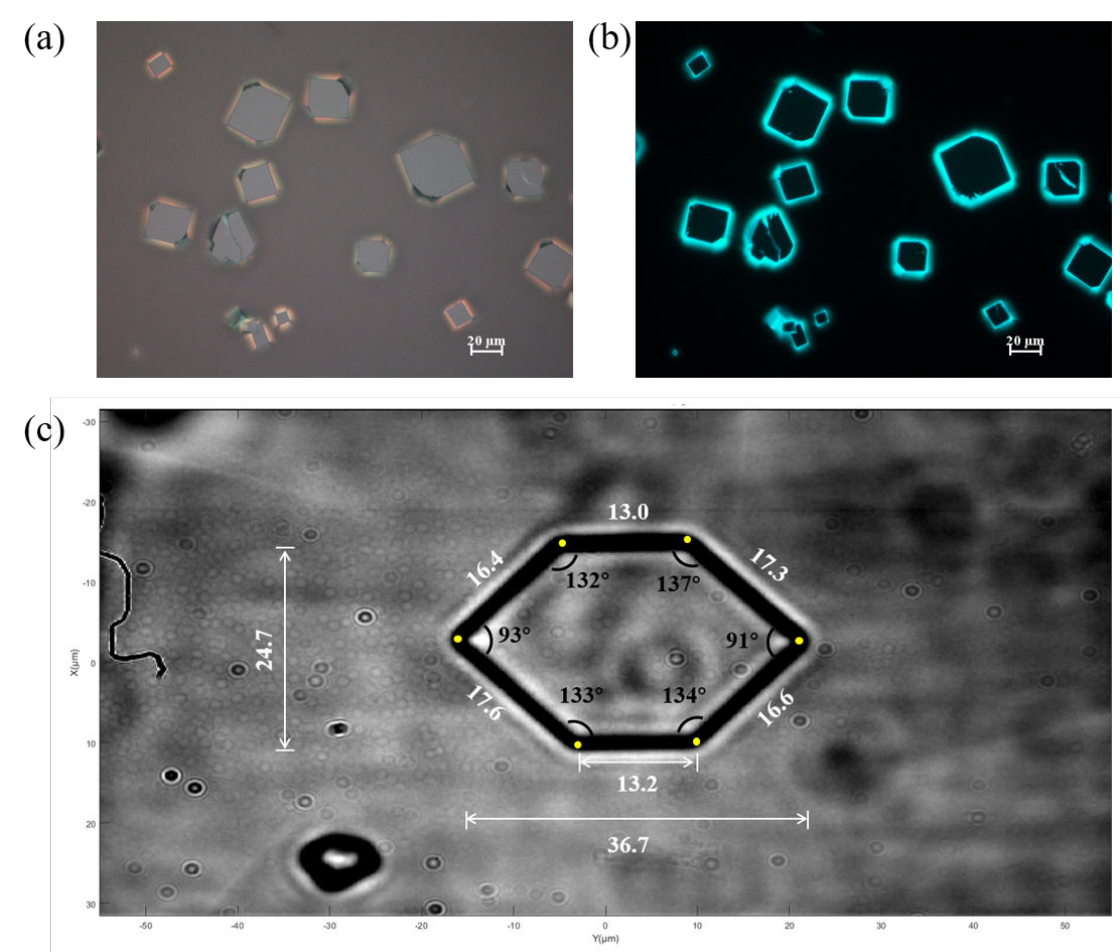}
    \caption{The bright (a) and dark (b) field images of BPDBNA hexagonal sheet. (c, e) Atomic force microscopy (AFM) image of as-prepared BPDBNA hexagonal sheets.}
    \label{figBPDBNA}
\end{figure}

Table S1. Crystal data and structure refinement for BPDBNA, 
Space Group	$P2_1/n$: 
\begin{center}
\begin{tabular}{c|c}
\hline
    $a$ & 54.1487~\AA \\
    $b$ & 6.71400~\AA \\
    $c$ & 7.2763~\AA \\
    $\alpha$ & 90.000~$^\circ$ \\
    $\beta$ & 92.496~$^\circ$ \\
    $\gamma$ & 90.000~$^\circ$ \\
\hline
\end{tabular}
\end{center}

\begin{figure}
    \centering
    \includegraphics[width=0.8\linewidth]{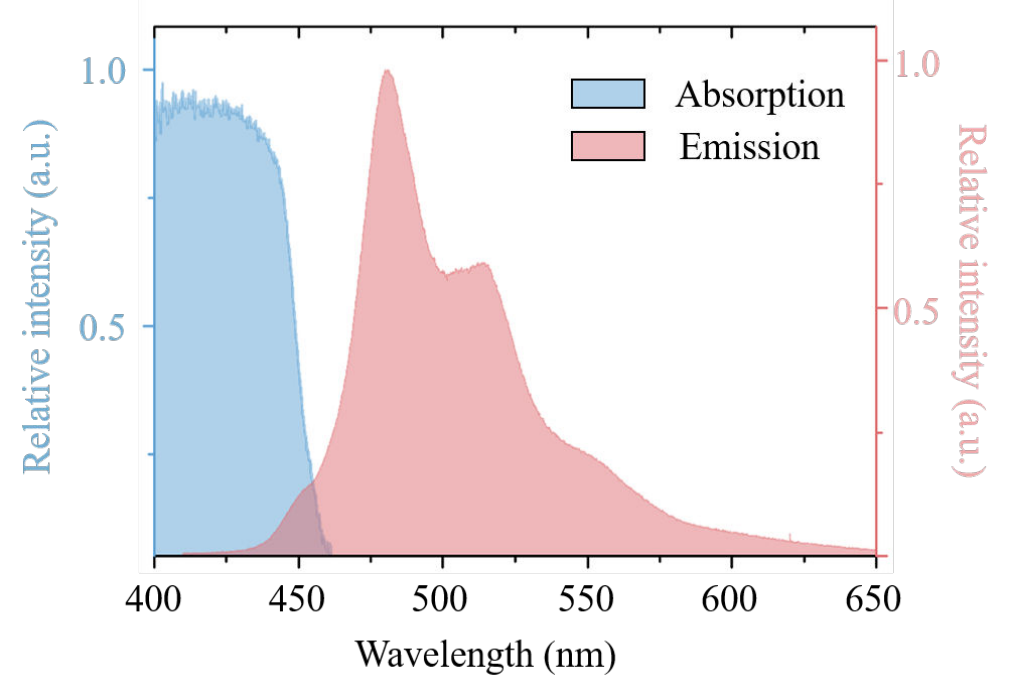}
    \caption{Absorption (pale blue area) and emission (pink area) spectra of BPDBNA hexagonal sheet.}
    \label{crystalAbsandPL}
\end{figure}

\begin{figure}
    \centering
    \includegraphics[width=1\linewidth]{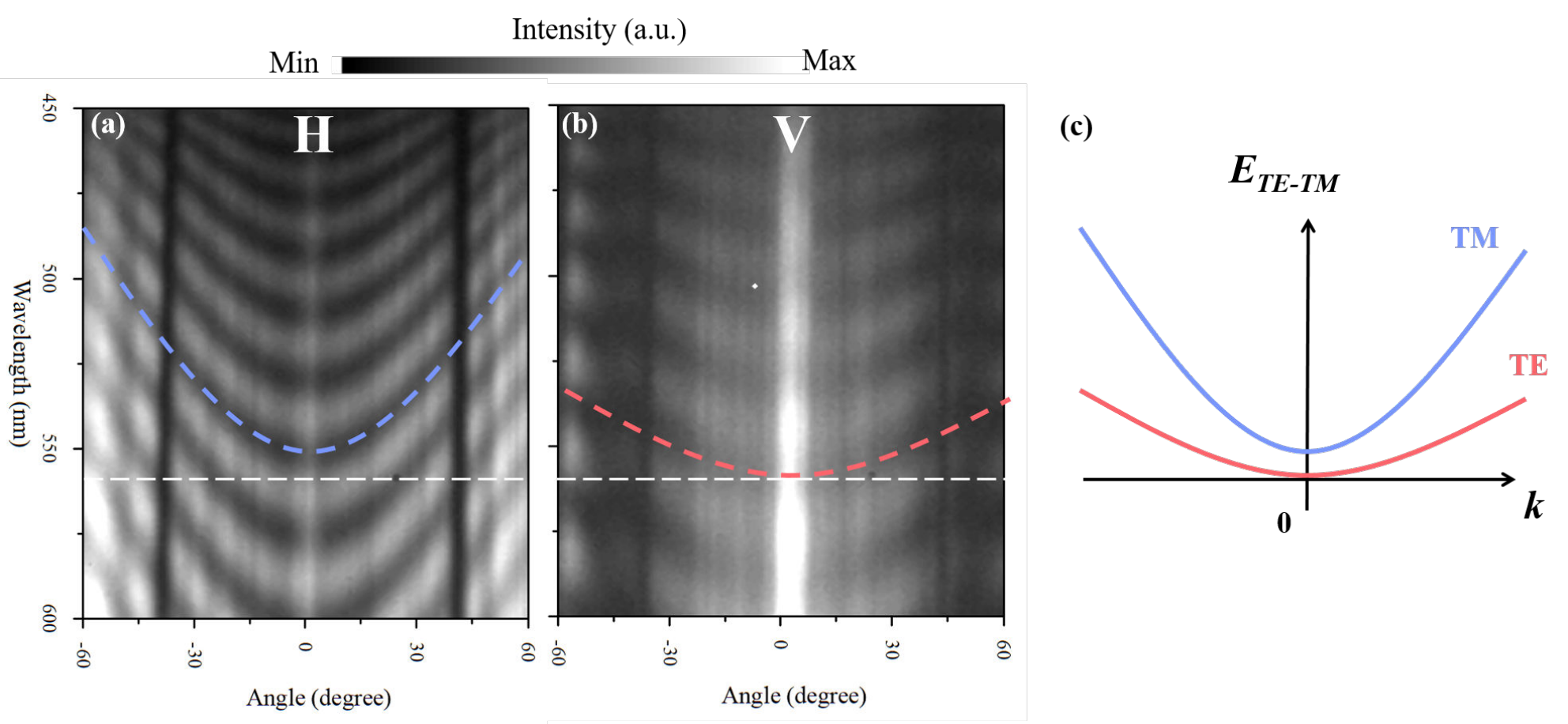}
    \caption{Measured k-space angle-resolved reflectivity spectra of a selected microcavity at room temperature. The ARR in horizontal (H) polarization (a) and vertical (V) polarization (b) along the direction perpendicular to the short side of the hexagonal crystal.(c) The schematic of TE and TM modes splitting.}
    \label{AbsARR}
\end{figure}

\begin{figure}
    \centering
    \includegraphics[width=1\linewidth]{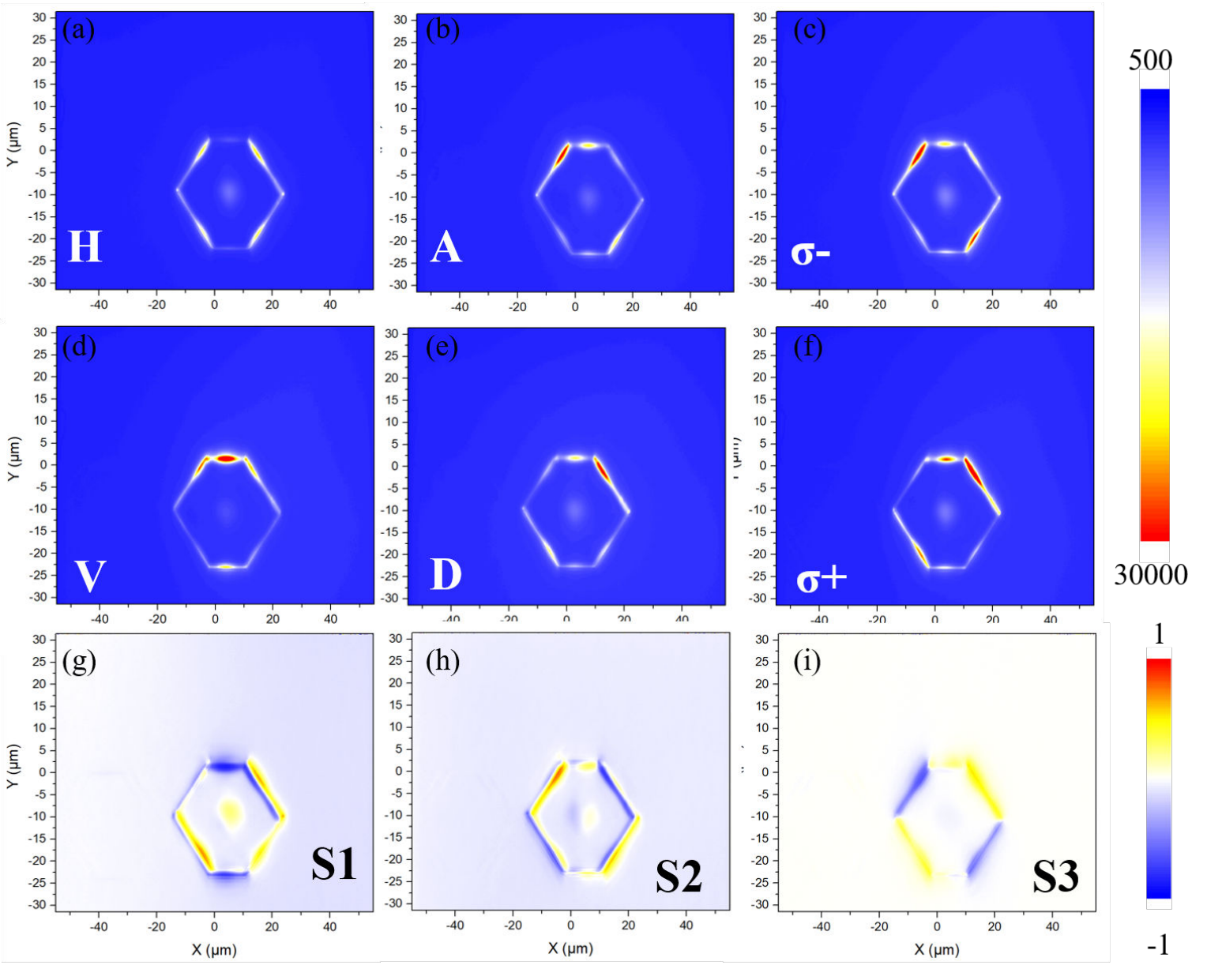}
    \caption{The PL of BPDBNA crystals in the 2D real space and its distribution of H(a), V(d), A(b), D(e), $\sigma$- (c) and $\sigma$+ (f) polarization. The distribution of the Stokes parameter S1(g), S2(h), S3(i).}
    \label{2DrealspaceofPL}
\end{figure}

\begin{figure}
    \centering
    \includegraphics[width=1\linewidth]{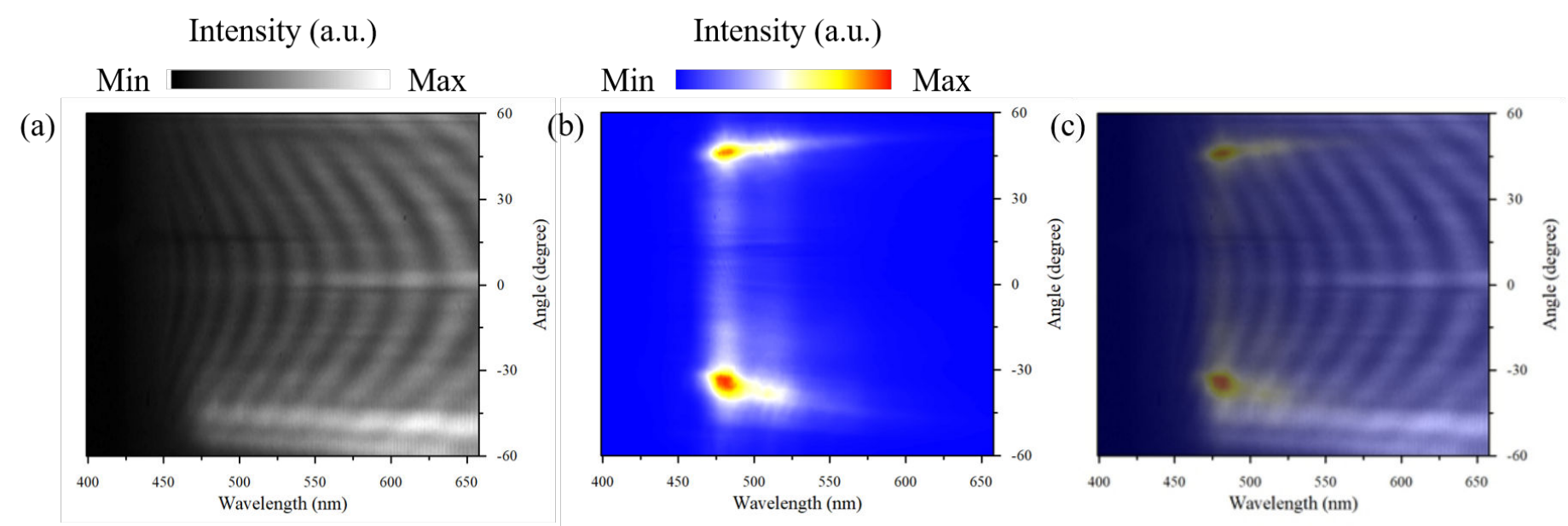}
    \caption{Angle-resolved reflection spectra of absorption (a) and luminescence (b), and their contrast (c).}
    \label{fig_PLARR}
\end{figure}

\begin{figure}
    \centering
    \includegraphics[width=1\linewidth]{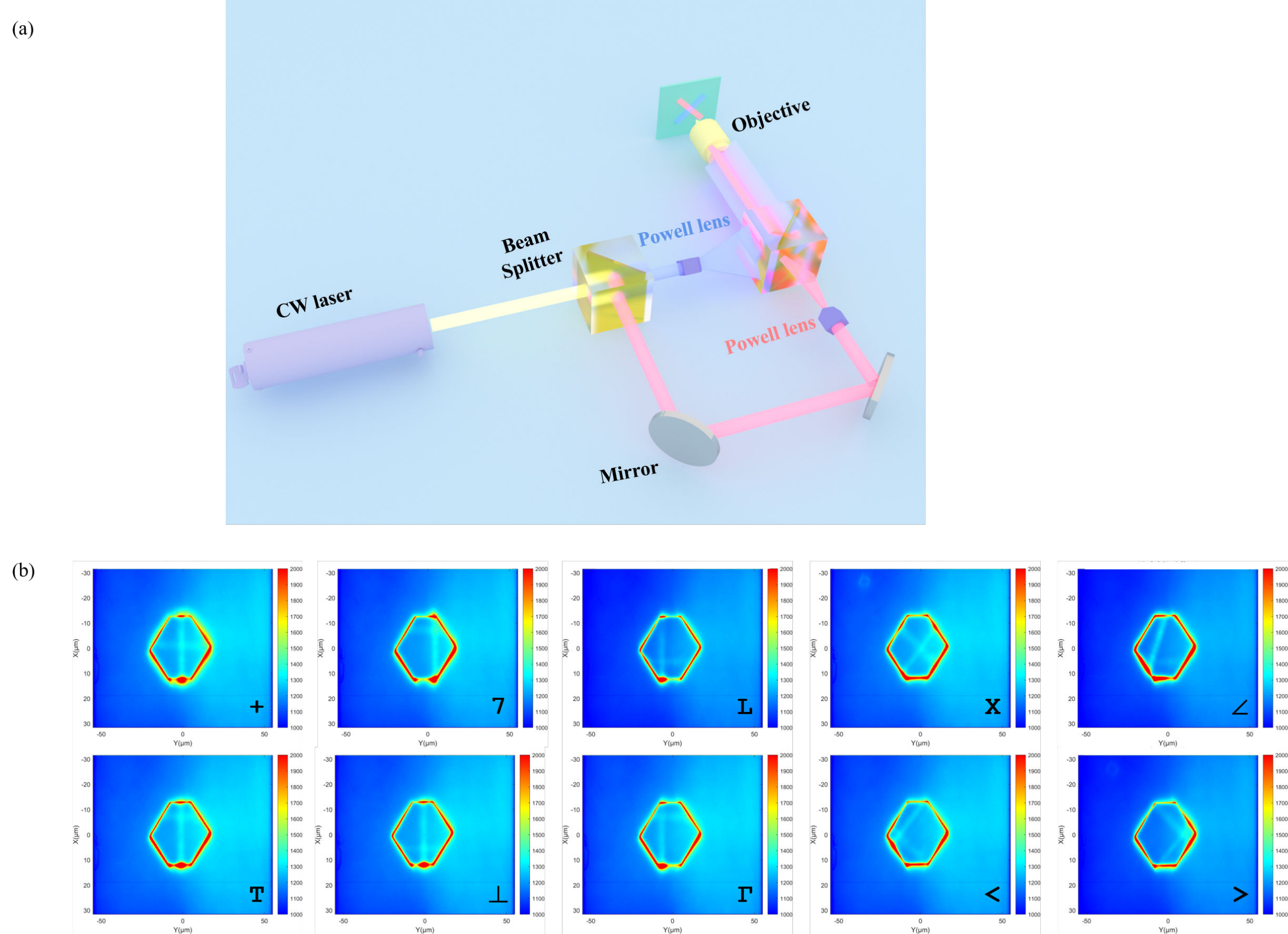}
    \caption{(a) Strip beam optical path diagram. (b)10 different patterns generated by strip beams.}
    \label{10symbols}
\end{figure}

\section{dataset collection}
\textbf{10 symbols dataset}
The excitation spot shapes of the ten symbols are achieved through the Powell Lens (Fig. S8). The 405 nm CW laser is split into two beams via a beam splitter. Subsequently, the Powell prism converts a point light spot into a strip beam. These two strip beams are then recombined via another beam splitter. By adjusting the direction and position of these two strip beams, ten distinct symbols can be achieved.

\begin{figure}
    \centering
    \includegraphics[width=1.0\linewidth]{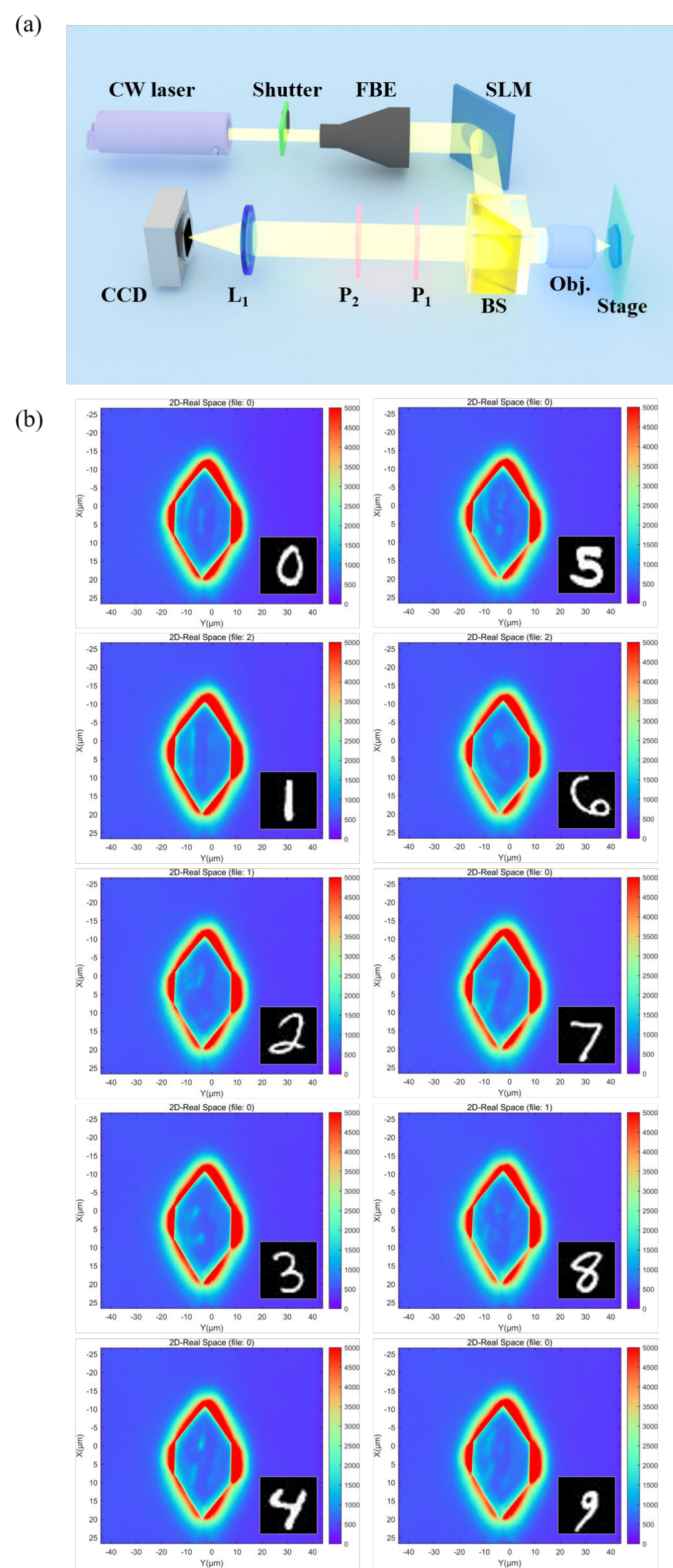}
    \caption{(a) Schematic diagram of the optical path for the MNIST experiment. (b)MNIST symbol-shaped CW Laser excitation sample.BS: beam splitter; L1: lenses; M1: mirror; SLM: Spatial Light Modulator; FBE: Fixed Magnification Achromatic Beam Expander; Obj.:objective lens; P1: Quarter-wave plate; P2: Linear polarizer. }
    \label{MNIST}
\end{figure}

\textbf{MNIST dataset}
By first passing the 405 nm CW laser through an FBE for beam amplification (Fig. S9), the size is optimally matched to the SLM (GAEA-2.1 Phase Only LCOS-SLM). The SLM generates light intensity shapes of MNIST digits, which are then vertically non-resonantly excited at the center position of the sample via a 100 X objective lens (NA = 0.95).By capturing PL of the sample using a CCD, single-shot imaging of the entire 2D real-space image is achieved. By setting a linear polarizer and quarter-wave plate on the detection optical path, 2D real-space images with different polarization states can be obtained. Finally, communication between the SLM and CCD is established via LabVIEW, enabling the automatic acquisition of 5000 digital data points. In our experiments, for the dataset comprising 35,000 images (6 sets of polarized states and unpolarized states corresponding to 5000 digits.), samples were collected over 20 hours (with 1-second acquisition intervals and 1-second gaps for data storage) The electric shutter is employed to ensure the excitation light remains switched off during data storage.

\textbf{Numerical simulations}
We describe the hexagonal resonator using the spinor Schrödinger equation, equivalent to the paraxial approximation of Maxwell's equations with polarization. This equivalence is based on the large quantization wave vector in the $z$ (growth) direction of the resonator, leading to the formation of a set of parabolic bands and determining their splittings and their effective masses (see Fig. 1(c) of the main text).
The equation is written on the circular basis $(\psi_+,\psi_-)^T$:
\begin{eqnarray}
i\hbar \frac{{\partial {\psi _ \pm }}}{{\partial t}} &=&  - \frac{{{\hbar ^2}}}{{2m}}\Delta {\psi _ \pm } - \beta {\left( {\frac{\partial }{{\partial x}} \mp i\frac{\partial }{{\partial y}}} \right)^2}{\psi _ \mp } - {\beta _0}{\psi _ \mp } + U{\psi _ \pm }\nonumber\\
&+& P(x,y,t) - \frac{{i\hbar }}{{2\tau }}{\psi _ \pm }
\end{eqnarray}
Here, $m$ is the effective mass of the quantized modes, $\beta$ is the TE-TM SOC strength, $\beta_0$ is the wavevector-independent splitting between linear polarizations due to the anisotropy of the BPDBNA organic material, $U$ is the hexagonal confining potential of the resonator, $\tau$ is the finite lifetime of the radiative modes, $\omega_0$ is the pumping laser frequency, and, finally, $P(x,y,t)$ is the linearly-polarized pump, whose spatial profile reproduces either the 10 simple symbols from the first dataset, or the MNIST digits (see Fig.~3(a) from the main text). This profile is obtained using an SLM in experiments. The time dependence of the pump corresponds to a white noise (spontaneous emission below the lasing threshold).

\begin{figure}
    \centering
    \includegraphics[width=0.95\linewidth]{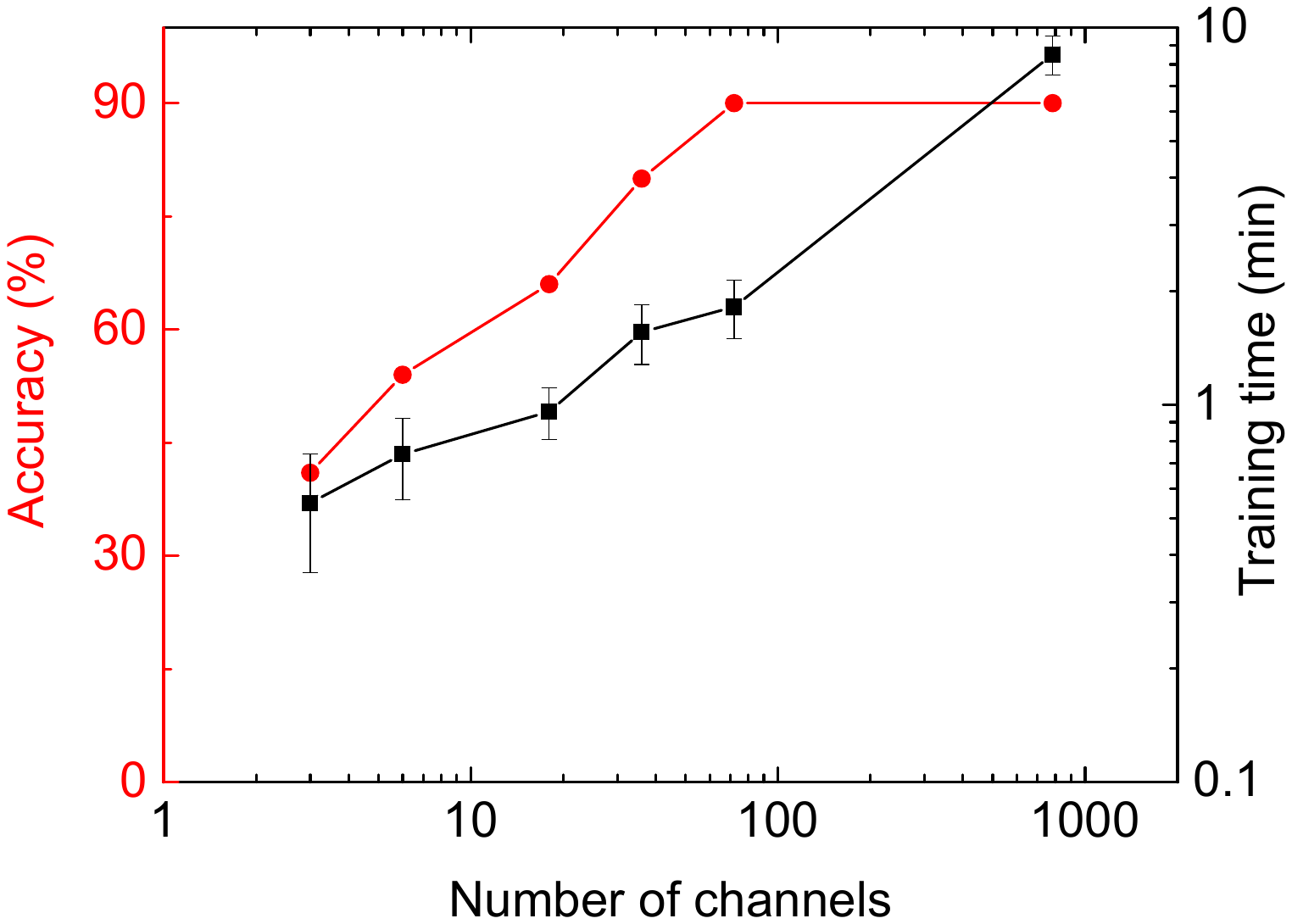}
    \caption{Accuracy (red) and the training time (black) versus the number of channels for shallow neural networks trained on the data coming from numerical simulations. The rightmost value corresponds to the original MNIST dataset resolution.}
    \label{figSupThMnist}
\end{figure}

The resulting wave functions $\psi_+$ and $\psi_-$ are used to calculate 
the spatial distribution of intensities in each of the 6 polarizations, which are then spatially selected at the edge of the sample and divided by sectors, as in experiment. The integrated intensities are then used to train the neural network. The results of the training for the 5000 symbols of the MNIST dataset are shown in Fig.~\ref{figSupThMnist}. Qualitatively, the behavior is very similar to the experiment. However, the achieved accuracy is actually lower than in the experiment, which may be explained by insufficient optimization: the simulations are actually taking much longer time than the experiment, which is another demonstration of the usefulness of the reservoir computing approach.

\end{document}